\documentclass[%
aps,
preprint,
 amsmath, amssymb]{revtex4-1}
\usepackage[english]{babel}
\usepackage{graphicx}
\usepackage{dcolumn}
\usepackage{bm}
\usepackage{textcomp}
\usepackage{color}
\usepackage{natbib}

\begin{document}
\title{Adatom-induced dislocation annihilation in epitaxial silicene}
\author{A. Fleurence} \email{Author to whom correspondence should be addressed: antoine@jaist.ac.jp}\affiliation{School of Materials Science, Japan Advanced Institute of Science and Technology, 1-1 Asahidai Ishikawa  923-1292, Japan}
\author{Y. Yamada-Takamura}\affiliation{School of Materials Science, Japan Advanced Institute of Science and Technology, 1-1 Asahidai Ishikawa  923-1292, Japan}

\begin{abstract}
The transformation of the stripe domain structure of spontaneously-formed epitaxial silicene on ZrB$_2$ thin film into a single-domain driven by the adsorption of a fraction of a monolayer of silicon was used to investigate how dislocations react and eventually annihilate in a two-dimensional honeycomb structure. The in-situ real time STM monitoring of the evolution of the domain structure after Si deposition revealed the mechanisms leading to the nucleation of a single-domain into a  domain structure  through a stepwise reaction of partial dislocations. After its nucleation, the single-domain extends by the propagation of edge dislocations at its frontiers. The identification of this particular nucleation-propagation formation of dislocation-free silicene sheet provides insights into how crystallographic defects can be healed in two-dimensional materials.
\end{abstract}
\maketitle
\date{\today}
 \section{Introduction}
Dislocations are linear irregularities in the ordering of crystals, which strongly influence their optical, electrical, chemical and mechanical properties. Although their presence can result from an imperfect crystallization, they also appear to reduce the strain in the crystals at their surfaces \cite{Harten85} or in epilayers \cite{Brune94,Lin04,Hwang95,Gunther95, Butz14}.  Dislocations are carrying a topological  charge, the Burgers vector, which is conserved upon  their migration or reaction. Such a reaction can be for instance the spontaneous decomposition of a normal dislocation, i.e. with a Burgers vector equal to a lattice vector, into partial dislocations, i.e. with  Burgers vectors whose sum is equal to the Burgers vector of the normal dislocation. Dislocation reactions can also be induced by applying external strain \cite{Schaff01, Wang10,Chauraud19} or  by depositing  adatoms \cite{Thurmer04,Figuera03}. \\ Silicene is a graphene-like two-dimensional material made of silicon atoms, which has been observed experimentally on a limited number of substrates \cite{Fleurence12,Vogt12, Chen12,Ming13,Aizawa14}. In contrast to graphene for which the robust sp$^2$ hybridization  of the orbitals in C atoms forces the covalent bonds to remain in the basal plane, the intermediate sp$^2$/sp$^3$ hybridization of the orbitals in Si atoms \cite{Cahangirov09} makes  the atomic structure of silicene buckled and highly flexible. The stripe domain structure consisting of an array of partial dislocations and texturing the epitaxial silicene sheet spontaneously forming on top of zirconium diboride (ZrB$_2$)  thin film grown on Si(111) \cite{Fleurence12,Fleurence17}  is a perfect test bench to study how dislocation  reaction occurs in  honeycomb structures.  Such a reaction can be induced by the deposition of a fraction of monolayer (ML)  of Si atoms  turns the  domain structure   into a single-domain  \cite{Fleurence16}.\\
  In this publication, we report the results of a   real-time scanning tunneling microscopy (STM) monitoring   of the evolution  of  this partial dislocation network after deposition of Si atoms. With the support of density functional theory (DFT) calculations, it revealed  the  non-trivial  reaction path  which, in contrast to  previously reported adatom-induced reactions of partial dislocations \cite{Thurmer04,Figuera03},   leads to their annihilation  and to the formation of a dislocation-free two-dimensional crystal.   
  \section{Experimental and computational details}
 ZrB$_2$(0001) thin films were grown on Si(111) by chemical vapor epitaxy in ultrahigh vacuum (UHV) \cite{YYT10}. The silicene layer was generated  by annealing the sample in a separate UHV system at 800\textdegree{}C for 30 mins which removes the natural oxide layer formed on ZrB$_2$ thin film upon transfer in air \cite{Fleurence12}. Silicon  was evaporated with a rate of 0.030 ML silicene per min (1ML silicene corresponds to 1.73 $ \times  10^{15}$  at.cm$^{-2}$) by direct current heating a silicon piece. Room temperature  STM observations were realized with a UHV-STM system (JSPM-4500A,JEOL) and STM images at higher temperature  were acquired with an  OMICRON variable temperature STM. Pt-Ir and W tips  were used respectively for the former and  latter systems. Otherwise specified,  constant-current STM images were recorded with a sample bias voltage of 500 mV and a tunnel current of 500 pA.\\
 DFT calculations within a generalized gradient approximation (GGA) \cite{Kohn65,Perdew96} were performed using the OPENMX code \cite{Ozaki03,Ozakiwebsite} based on norm-conserving pseudopotentials generated with multireference energies \cite{Perdew96} and optimized pseudoatomic basis functions.  The domain structure of  epitaxial silicene on ZrB$_2$(0001) thin film was modeled by ZrB$_2$ slabs consisting of two Zr, one B, and one silicene layers. A 32 \AA-thick vacuum space was used.  The structure of silicene was compared to that of a silicene layer put in contact with a thicker ZrB$_2$ slab, to make sure that the effect of the slab ultimate thickness is not significant. A ZrB$_2$ slab without silicene was also considered, as reference. For Zr atoms, three, two, and two optimized radial functions were allocated for the s, p, and d orbitals, respectively, as denoted by s3p2d2. For both Si and B atoms, s2p2d1 basis functions were adopted. A cutoff radius of 7 Bohr was chosen for all the basis functions. A regular mesh of 150 Ry in real space was used for the numerical integrations and for the solution of the Poisson equation together with a (1 $\times$ 3 $\times$ 1) mesh of the  reciprocal space. For the geometrical optimization, the bottom Zr layer is fixed and the other atoms can relax freely.  The force on each atom was relaxed to be less than 0.0001 Hartree/Bohr. Calculations were performed in the `4 domains supercell' to account for the domain structure of the silicene sheet.  Details of the structures considered for this study can be found in the supplementary materials \cite{SM}.

\section{Partial dislocations  in epitaxial silicene on ZrB$_2$(0001)}
\begin{figure}[h!]
	\centering
	\includegraphics[width=7 cm]{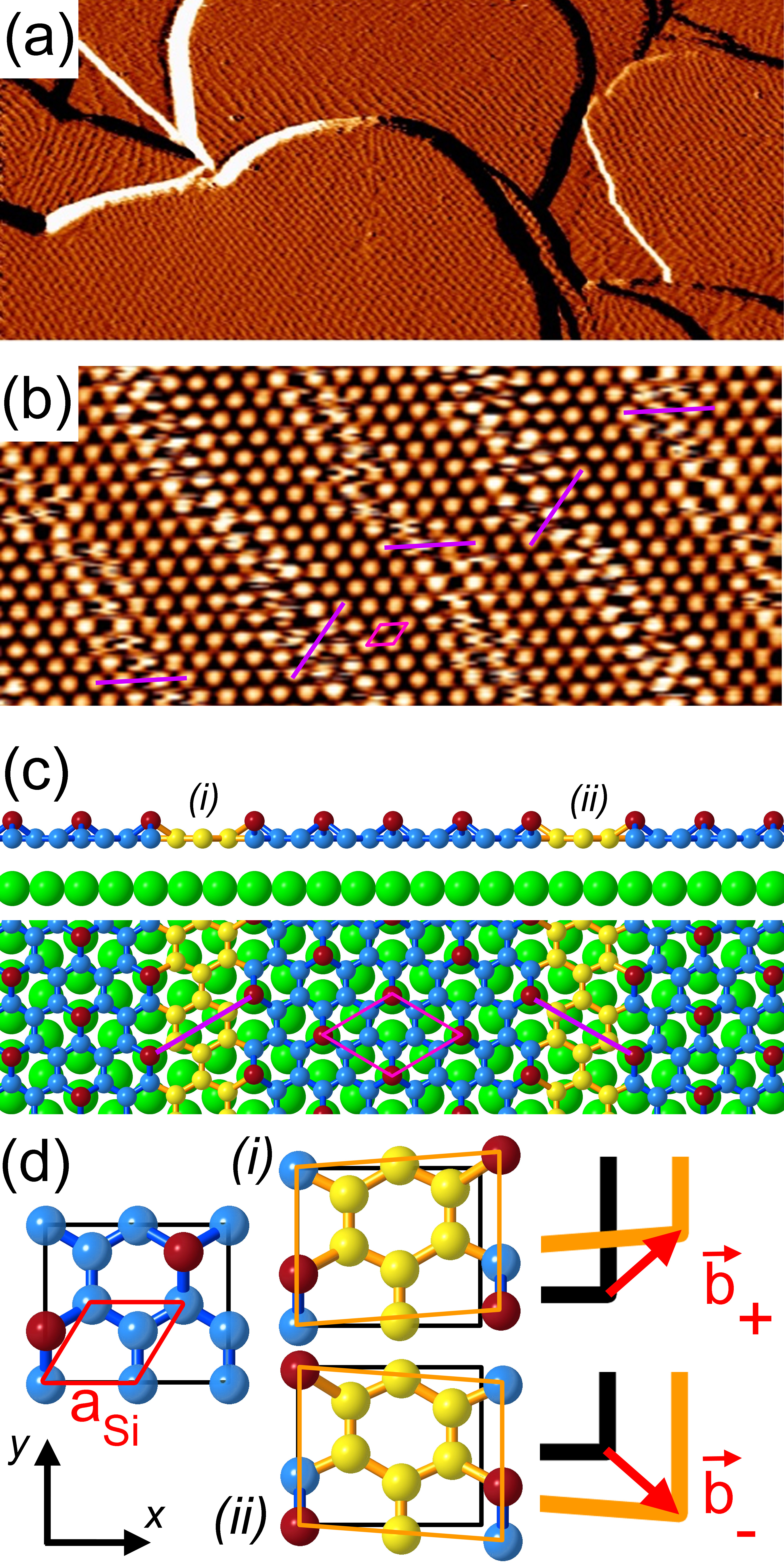}
\caption{\textbf{Domain structure of epitaxial silicene}. (a) Differentiated STM image ($V$=-300 mV, $I$=61 pA, 300 nm $\times$ 130 nm) of the silicene sheet covering the ZrB$_2$(0001) thin film surface. (b): STM image ($V$=-800 mV, $I$=400 pA, 20 nm $\times$ 9 nm) of the periodic  domain structure. (c): Top- and side views of a stick and ball model of epitaxial silicene. Zr atoms are in green.  Protruding, bottom, and boundaries atoms are respectively red-, blue- and yellow-colored. The  ($\sqrt{3}\times\sqrt{3}$) unit cell is indicated by rhombus. The purple lines emphasize the direction of the shifts between protrusion arrays (d): Honeycomb structure of silicene in the domains (blue) and at the boundaries (yellow). The unitcell of unreconstructed silicene is indicated by a red rhombus. The Burger vectors of the dislocations are indicated in red. The x and y directions correspond respectively to ZrB$_2[1\overline{1}00]$ and ZrB$_2[11\overline{2}0]$.} 
\label{Fig1}
\end{figure}

  The differentiated large-scale STM image of Fig. \ref{Fig1}.(a) shows the silicene sheet spontaneously crystallized on ZrB$_2$(0001), which is textured by  one-dimensional periodic domain structures  having a pitch of 2.7 nm along  $<1\overline{1}00>$ directions of ZrB$_2$. The spontaneous formation of the domain structures  originates from the  5\% compressive epitaxial strain which forces the silicene structure to adopt a lattice parameter $a_{Si}$ of 3.65 \AA\enspace inside the domains     to  make the ($\sqrt{3}\times\sqrt{3}$) unit cell of silicene fit the ($2\times2$) unit cell of ZrB$_2$(0001).  As shown in Fig. \ref{Fig1}.(c), the Si atoms in the  ($\sqrt{3}\times\sqrt{3}$)-reconstructed silicene unit cell   are located 2.3 \AA\enspace  above the Zr top layer except that sitting above a Zr atom, which is protruding  at 3.9 \AA\enspace \cite{Lee13,Lee14ARPES}.  The epitaxial strain is partially released by the formation of partial dislocation, namely domain boundaries which forms a period one-dimensional structure due to the repulsion between domain boundaries through the stress field they generate \cite{Nogami19}.  As seen in Fig. \ref{Fig1}.(c), the partial dislocations  are $3\sqrt{3} a_{ZrB_2}/2$-wide (where  $a_{ZrB_2}$=3.18 \AA\enspace is the lattice parameter of the ZrB$_2$(0001))    and are of two different types, symmetric with respect to each other in a mirror symmetry operation and perfectly alternated along the direction perpendicular to the domains.  Their respective Burgers vectors are $\vec{b}_\pm=\frac{a_{Si}}{4}(\vec{u}_x \pm \frac{1}{\sqrt{3}}\vec{u}_y) $ where $\vec{u}_x$  and $\vec{u}_y$ are the unitary vectors  respectively along the directions perpendicular and parallel to the domain boundaries.   This  gives rise to 4 equivalent domains with  sequential shifts of the (unreconstructed) Si honeycomb structure, labelled  $a$, $b$, $c$, $d$ in the following (See Fig. \ref{Fig2}.(a)). It has to be noted that the atomic structure of epitaxial silicene in domains $a$ and $c$ (or $b$ and $d$) can be permuted by a translation of $\frac{a_{Si}}{2}\vec{u}_x$, whereas the permutation of the atomic structures  in domains  $a$ and $b$ (or $b$ and $c$)  involves a rotation by an angle of 180 \textdegree{}.
   As a matter of fact, these 4 shifts of the silicene honeycomb structure correspond to 4 different shifts of the array of  protruding atoms.    As a consequence, the resolution of the protrusions in STM images and the observation of the shift direction between  neighboring domains  allow for determining the positions of all Si atoms in the silicene sheet. Note that owing to different local densities of states in the domains and in the boundaries \cite{Warner17}, the domain structures can be easily visualized by choosing a proper bias voltage, even in large scale images such as the one in Fig. \ref{Fig1}.(a).
   \section{Nucleation of a single-domain by Si adatoms-induced reaction of partial dislocations}
\begin{figure}[h!]
	\centering
		\includegraphics[width=16 cm]{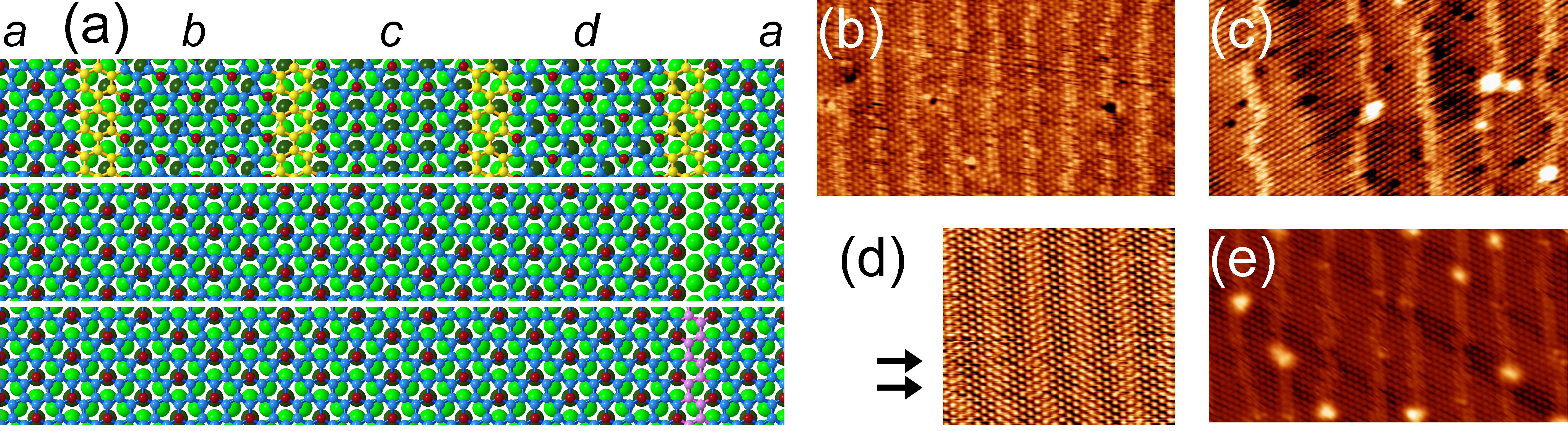}
\caption{\textbf{From domain-structure to single-domain}. (a): Schematics of the transformation of the stripe domains into a single-domain by the reaction of four partial dislocations followed by the incorporation of Si atoms (colored in pink). (b) and (c): STM images (30 nm $\times$ 16 nm) of the domain structure before and after the deposition of 0.015 ML Si at 350 \textdegree{}C. (d): STM image  (20 nm $\times$ 16 nm) recorded at 300  \textdegree{}C. The arrows are pointing to the place where collective motion of the boundaries occurred during the scan. (e): STM image after deposition of 0.015 ML of Si atoms at RT.}
\label{Fig2}
\end{figure}

\begin{figure}[h!]
	\centering
		\includegraphics[width=16 cm]{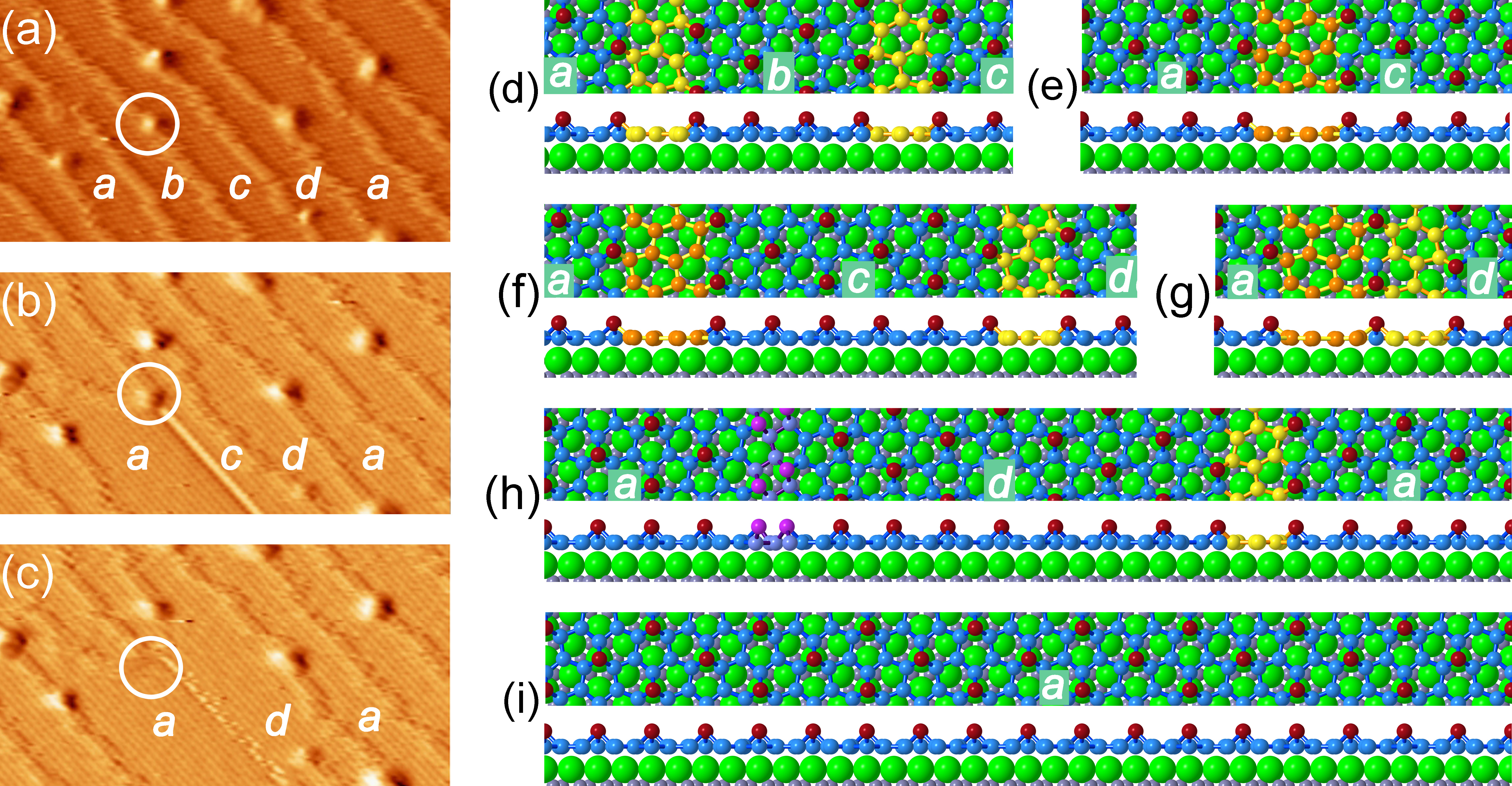}
\caption{\textbf{Nucleation of a single-domain in epitaxial silicene by the reaction of partial   dislocations}.  (a)-(c): Differentiated STM  images (40 nm $\times$ 20 nm) recorded 996, 1253, and 1270 minutes after the Si deposition. (d)-(g):  Top and side views of optimized structures  for   $abc$,  $ac$, $acd$, $ad$  domains sequence with 104 Si atoms in the 4 domains unit cell. (h) and (i) $ada$ domain sequence and $a$ single-domain with 108 Si atoms in the 4 domains unit cell.}
\label{Fig3}
\end{figure}

As depicted in Fig. \ref{Fig2}.(a), the transformation of the  $abcda$ domain sequence of the pristine silicene layer into a single-domain $a$ needs the reaction of four partial dislocations  giving rise to a normal  dislocation (with a perfect Burgers vector of $2\vec{b}_+ +2\vec{b}_-=a_{Si}\vec{u}_x$)  permitting the integration of silicon atoms into the silicene lattice. The creation of such a normal dislocation   would represent  a cost of energy of about 16 meV  per Si atom of the  4 domains supercell according to DFT calculations (See Table \ref{Table} and Figure 1 in Supplementary materials \cite{SM}). It would involve the unlikely coherent motion of a large number of Si atoms that would need to overcome the repulsion between domain boundaries. The collective motion of the domain boundaries pointed out by arrows in the STM image of Fig. \ref{Fig2}.(d) shows  that this interaction  is also effective in the substrate temperature range (300-350\textdegree{}C) at which the formation of single-domain  can be realized by Si deposition  \cite{Fleurence16}. Intriguingly, as reported in Table \ref{Table}, the formation of single-domain corresponds  to an even higher increase of energy (30 meV per Si atoms of the 4 domains supercell), which points out that the adsorption  of Si atoms on silicene sheet must represent  a cost of energy even higher to make the reduction of the number of  Si adatoms a driving force for the transformation of the domain structure into a single-domain. This gain of energy was evaluated by considering  domain-structures on which Si atoms  were  adsorbed above the center  of the domains  and above the center of the  boundaries, respectively. The formation of a single-domain from each of these structures corresponds respectively to a reduction of energy of 24.5 meV  and of 26.4 meV per Si atom of the supercell (See Table \ref{Table} and Figure 1 of the Supplementary materials \cite{SM}). It has to be noted that the adsorbtion in the center of the domains is more favorable than in the center of the boundaries is in disagreement with the experimental observation. This is possibly due to an irrelevant choice for the adsorption sites above the boundaries and to the difference of energy between individual adatoms and Si clusters.\\
  As the time scale of the formation of single-domain silicene upon deposition on heated samples  is not ideal for real-time STM observations \cite{Fleurence16}, a different approach was considered to identify the nontrivial dislocation dynamics process leading to the silicene sheet transformation. To determine how the partial dislocations react to eventually incorporate the Si adatoms in the silicene lattice,  a small amount of Si atoms (0.015 ML) corresponding coarsely to half of the coverage required to turn the stripe domain structure into single-domain \cite{Fleurence16}  was deposited on epitaxial silicene at room temperature (RT), and the evolution of the silicene sheet during several hours after this deposition was monitored by STM. \\
  The flexibility of the atomic structure of silicene is exemplified by the fluctuation of the domain boundaries by $\sqrt{3}a_{ZrB_2}$ observed in STM images recorded at RT such as that of Fig. \ref{Fig1}.(b). As shown in Fig. S2.(b) of supplementary materials \cite{SM}, this motion of the boundaries result from the displacement of  a pair of atoms by a distance corresponding to the Burgers vector of the dislocation.  This fluctuation indicates that the energy barrier between the two equilibrium positions of these pair of atoms is shallow enough to allow for the thermal activation of the  free motion of the dislocations at RT in silicene, which is strikingly different from graphene whose robust bonding requires a much higher temperature \cite{Gong15} or  irridation \cite{Lehtinen13}. The reaction of the dislocations is therefore only limited by the repulsive interaction between the domain boundaries.\\  
The STM images of Figs. \ref{Fig2}.(c) and (e) compare the domain structures of silicene after Si depositions at 350 \textdegree{C}  and  at  RT. In the former case, the density of domain boundaries is reduced by half while their ordering  is locally preserved which suggests that the interaction between them is still acting for such larger periodicities. In contrast to the case of heated samples, the deposition at RT did not cause the vanishing of domain structures except at parse locations, but gives rise to Si clusters with  an apparent height of 1 \AA,  an apparent size of 2-3 nm, an estimated density of  10$^4 \mu m^{-2}$  and a first neighbor distance of 6 nm. The clusters  are sitting systematically  at the domain boundaries.\\
The sequence of STM images shown in Figs. \ref{Fig3}.(a)-(c) presents the evolution of the silicene sheet in the presence of these Si clusters. The Fig. \ref{Fig3}.(a)  depicts a $abcda$ domain sequence similar to that of the pristine domain structure. The STM image of Fig. \ref{Fig3}.(b) recorded 257 mins after that of Fig. \ref{Fig3}.(a),  shows that along a distance of 20 nm  a pair of domain boundaries $a/b$ and $b/c$ disappeared  together with the  domain $b$.  The so created $a/c$ boundary, which sits near  the location of the $b/c$ boundary, consists of a straight 11 \AA-wide gap between the protrusions rows of the $a$ and $c$ domains. 
This feature is consistent with the ($2\sqrt{3}a_{ZrB_2}$)-wide dislocation having a Burgers vector of $\vec{b}_{a/c}=\frac{a_{Si}}{2}\vec{u}_x$, whose optimized structure is shown in  Fig. \ref{Fig2}.(e).  Since $\vec{b}_{a/c} =\vec{b}_+ +\vec{b}_-$, this $a/c$ boundary can be identified as the result of the  reaction of  partial dislocations  $a/b+b/c \rightarrow a/c$. The fact that this type of partial dislocation is never observed in the domain structure of pristine silicene is consistent with the fact that this reaction is  in violation of the  Frank criterion which predicts that such $a/c$ dislocation must decompose into $a/b$ and $b/c$ dislocations as $\|\vec{b}_{a/c} \|^2>\|\vec{b}_{a/b}\|^2+\|\vec{b}_{b/c}\|^2$.  DFT calculations confirm that  such a reaction  represents an increase in energy of the silicene sheet by about  4 meV per Si atoms of the 4 domains supercell (See Table \ref{Table}).   As the reaction does not involve the incorporation of any Si atoms in the silicene sheet, its spontaneous  evolution towards a metastable structure in spite of the repulsion between domain boundaries  can only be explained by the modification of the stress field caused by the clusters. One can observe that the  position of this $a/c$ boundary does not fluctuate much  in comparison with the partial dislocations with $\vec{b}_{\pm}$ Burgers vectors. This can be explained by the fact that the shift of this boundary by a  distance of $\sqrt{3}a_{ZrB_2}$   needs the motion of 3 atoms  (Fig. S2.(b) in supplementary materials) instead of  two in the case of dislocations with $\vec{b}_{\pm}$ Burgers vectors. One can deduce that the energy barrier between the two positions is high enough to limit the thermal activation of the gliding of this dislocation at RT. \\
The domain structure continues to evolve after this first reaction as around 27 min after the moment the image of Fig. \ref{Fig3}.(b) was recorded, the $acda$  domains sequence turned  into a $ada$ sequence.  The $a/d$ domain boundary appears in the STM image as a narrow and fuzzy 9 \AA-wide   feature.   A simple  $a/c+c/d \rightarrow a/d$  reaction would  give rise to a dislocation with a Burgers vector of $\vec{b}_{a/d} =\vec{b}_+ +2\vec{b}_-=\frac{a_{Si}}{4}(3\vec{u}_x-\frac{1}{\sqrt3}\vec{u}_y)$. This possibility was evaluated  in DFT calculations by considering   a dislocation spanning a width of $\frac{3}{2}\sqrt{3}a_{ZrB_2}$. Upon geometrical optimization, this dislocation decomposed  in a pair  of $a/c$ and $c/d$ dislocations  sitting side-by-side (Fig. \ref{Fig3}.(g)) resulting in an  overall width of the $a/d$ boundary inconsistent with that observed in STM image of Fig. \ref{Fig3}.(c). It has to be noted that the comparison of the energies (Table \ref{Table})  of the structures of Figs.  \ref{Fig3}.(f) and (g) indicates that the  closer proximity of these two dislocations represents an increase of energy which reflects the repulsive interaction between boundaries. \\
  The comparison of the images of Figs. \ref{Fig3}.(b) and (c) indicates that the dislocation reaction is accompanied by the vanishing of the Si cluster marked by a white circle. One can deduce that silicon atoms were incorporated in  the silicene lattice at this stage of the silicene sheet transformation. To verify this hypothesis, a  supercell in which a row of Si atoms was added between the domains $a$ and $d$ was considered. This domain boundary is therefore a dislocation with a Burgers vector of   $\vec{b}_{a/d} =\vec{b}_{a/c}+\vec{b}_{c/d}-a_{Si}\vec{u}_x= -\frac{a_{Si}}{4}(\vec{u}_x+\frac{1}{\sqrt3}\vec{u}_y)=-\vec{b}_{+}$. The optimized structure  visible in Fig. \ref{Fig3}.(h) suggests that this boundary is made of highly  compressed hexagon rings along which  protruding atoms are separated by a single row of Zr atoms. In contrast to the structure of Fig. \ref{Fig3}.(g), this structure  is consistent with the boundary observed in the STM image shown in Fig. \ref{Fig3}.(c). The apparent width of the  boundary can be explained by the fluctuation of the pair of Si atoms marked by a red circle in Fig. S2.(c) of supplementary materials \cite{SM} allowed  by an energy barrier even shallower than for the boundaries with Burgers vectors of $\vec{b}_\pm$.\\
   According to DFT calculations, the formation of the $ada$ sequence with the incorporation of Si atoms, represents a much higher cost of energy than the simple $a/c+c/d \rightarrow a/d$ reaction. The silicene sheet must transit by the structure shown on Fig. \ref{Fig3}.(f), which may explain the position of the $a/d$ boundary slightly on the side of the position of the $a/c$  boundary. \\
 However, one may conceive the following explanation based on the fact that the position of the $a/d$ dislocation is the same as that of the $a/c$ dislocation. The tension  at the $a/c$ boundary weakens the silicon bonds and the honeycomb lattice eventually cracks. As a consequence of it, the domain $c$ is pulled to the left to annihilate the $c/d$ dislocation.  On the other side, the so-created empty space separating dangling bonds can be filled by Si atoms which can crystallise in the form of an honeycomb structure in spite of the compressive strain exerted on it.\\ 
 After the record of the STM image in Fig. \ref{Fig3}.(c), no further evolution of the surface was observed until the tip eventually degraded. One can easily  conceive that the $ada$ domain sequence  could have evolved into a single  $a$ domain (or a single $d$ domain) by reaction of the $a/d$ dislocation with  the $d/a$ dislocation on its right (or the $d/a$ dislocation on its left). Their annihilation  ($\vec{b}_{a/d}+\vec{b}_{d/a}=0$) would  reduce the energy of the silicene sheet by 9 meV per Si atom of the 4 domains supercell (Table \ref{Table}) and would lead to the nucleation of a single-domain as the one whose optimized structure is shown in Fig. \ref{Fig3}.(i) and observed experimentally in Fig. \ref{Fig4}. Please note that the single-domain imaged in this figure nucleated before the dislocations reaction shown in Fig. \ref{Fig3}, which indicates that single-domains keep nucleating while already formed single-domains expand.
\section{Propagation of a  single-domain}
\begin{figure}[h!]
	\centering
		\includegraphics[width=16 cm]{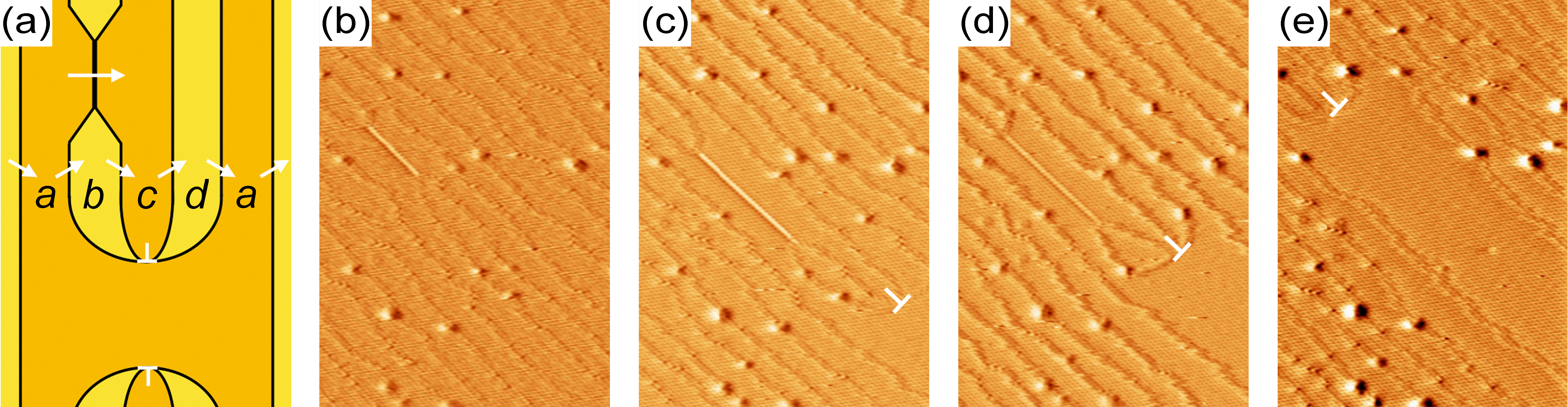}
\caption{\textbf{Propagation of single-domain}. (a): Schematics of single-domain nucleated within the domain structure. Burgers vector of  partial dislocations and the edge dislocations at the frontier are indicated (b)-(e): Differentiated STM images  (42 nm $\times$ 63 nm) recorded after 33, 198, 232 and 368 minutes after the Si deposition. The moving edge dislocation is indicated.}
\label{Fig4}
\end{figure}

As depicted in  Fig. \ref{Fig4}.(a), the nucleation of a 5 domains-wide single-domain in the domain structure results in interfaces with the domain-structure  where dislocations merge into normal edge dislocations. Note that such edge dislocations are also observed in the pristine silicene sheet at the frontiers between domain-structures with different orientations \cite{Nogami19}. However,  in contrast to  pristine silicene, the edge dislocations formed by the nucleation of a single-domain within a domain structure  propagate in such a way that the single-domain extends, as shown in  the STM images of Figs. \ref{Fig4}.(b)-(e) recorded at different times after the Si deposition.  Although this propagation is  limited by its highly anisotropic character allows for enlarging the single-domain much more rapidly than by the nucleation of new single-domains. This is exemplified  in the STM images of Figs. \ref{Fig4}.(b)-(d) in which  dislocations started reacting but the passage of the edge dislocation occurs before completion of the transformation into a single-domain.   The migration of the edge dislocation occurs mostly by climb steps resulting from the integration of Si adatoms  into the silicene sheet.  This is in contrast  with previous observation of the diffusion of dislocations in graphene for which the climbs are mostly  the result of the evaporation of  atoms   \cite{Gong15,Warner12,Lehtinen13}. It has to be noted that the extension of the single-domain is  made possible  by a diffusion length at RT large enough to allow for the Si adatoms from the surrounding clusters to reach the edge dislocation.The propagation of the single-domain gives rise to the vanishing of all Si clusters on the passage of the interface between single-domain and the stripe domain structure.  The comparison of the  clusters surrounding the single-domain in Figs. \ref{Fig4}.(b) and (e)  points out a slight increase of their  size  due to the attachment of Si atoms which were not incorporated into the single-domain. These enlarged Si islands may lose their mobility and therefore hinder the nucleation of new single-domain. This may explain why the surface stops evolving towards larger single-domain area. At temperatures as high as 300\textdegree C \cite{Fleurence16}, this effect may not occur and the silicene sheet  fully transforms into a single-domain.
\section{Conclusions}
By means of a real-time STM monitoring, we experimentally determined the complex nucleation-propagation process through which an array of partial dislocations in epitaxial silicene on ZrB$_2$(0001) turns into a single-domain  in presence of Si adatoms.  This transformation involves first the  nucleation of a single-domain through a sequence of  reaction of partial dislocations and incorporation of adatoms into the silicene sheet. Noteworthy, the presence of adatoms on top of silicene initiates the reaction of dislocations even before they get integrated into the silicene sheet.  The single-domain then extends anisotropically by the  motion of edge dislocations leading to further  incorporation of Si adatoms.  Beyond  insightful information on the dynamics of dislocations in two-dimensional honeycomb structure,   these experimental observations indicate  how dislocations can be removed from crystals in spite of their stability. It may thus help finding solution to heal crystallographic   defects  in surfaces, interfaces, 2D materials, or any kind of nanomaterials.
\section{Acknowlegments}
A.F. acknowledges financial support from JSPS KAKENHI Grant Number 26790005. We sincerely thank Prof. T. Ozaki and Dr C.-C. Lee of The University of Tokyo for their help in carrying out DFT calculations. We are grateful to Prof. C. Coupeau  of the University of Poitiers for fruitful discussions and his kind reading of our manuscript.
\section{Competing interests}
The authors declare no competing interests.
\newpage


\begin{thebibliography}{32}%
\makeatletter
\providecommand \@ifxundefined [1]{%
 \@ifx{#1\undefined}
}%
\providecommand \@ifnum [1]{%
 \ifnum #1\expandafter \@firstoftwo
 \else \expandafter \@secondoftwo
 \fi
}%
\providecommand \@ifx [1]{%
 \ifx #1\expandafter \@firstoftwo
 \else \expandafter \@secondoftwo
 \fi
}%
\providecommand \natexlab [1]{#1}%
\providecommand \enquote  [1]{``#1''}%
\providecommand \bibnamefont  [1]{#1}%
\providecommand \bibfnamefont [1]{#1}%
\providecommand \citenamefont [1]{#1}%
\providecommand \href@noop [0]{\@secondoftwo}%
\providecommand \href [0]{\begingroup \@sanitize@url \@href}%
\providecommand \@href[1]{\@@startlink{#1}\@@href}%
\providecommand \@@href[1]{\endgroup#1\@@endlink}%
\providecommand \@sanitize@url [0]{\catcode `\\12\catcode `\$12\catcode
  `\&12\catcode `\#12\catcode `\^12\catcode `\_12\catcode `\%12\relax}%
\providecommand \@@startlink[1]{}%
\providecommand \@@endlink[0]{}%
\providecommand \url  [0]{\begingroup\@sanitize@url \@url }%
\providecommand \@url [1]{\endgroup\@href {#1}{\urlprefix }}%
\providecommand \urlprefix  [0]{URL }%
\providecommand \Eprint [0]{\href }%
\providecommand \doibase [0]{http://dx.doi.org/}%
\providecommand \selectlanguage [0]{\@gobble}%
\providecommand \bibinfo  [0]{\@secondoftwo}%
\providecommand \bibfield  [0]{\@secondoftwo}%
\providecommand \translation [1]{[#1]}%
\providecommand \BibitemOpen [0]{}%
\providecommand \bibitemStop [0]{}%
\providecommand \bibitemNoStop [0]{.\EOS\space}%
\providecommand \EOS [0]{\spacefactor3000\relax}%
\providecommand \BibitemShut  [1]{\csname bibitem#1\endcsname}%
\let\auto@bib@innerbib\@empty
\bibitem [{\citenamefont {Harten}\ \emph {et~al.}(1985)\citenamefont {Harten},
  \citenamefont {Lahee}, \citenamefont {Toennies},\ and\ \citenamefont
  {W\"oll}}]{Harten85}%
  \BibitemOpen
  \bibfield  {author} {\bibinfo {author} {\bibfnamefont {U.}~\bibnamefont
  {Harten}}, \bibinfo {author} {\bibfnamefont {A.~M.}\ \bibnamefont {Lahee}},
  \bibinfo {author} {\bibfnamefont {J.~P.}\ \bibnamefont {Toennies}}, \ and\
  \bibinfo {author} {\bibfnamefont {C.}~\bibnamefont {W\"oll}},\ }\href
  {\doibase 10.1103/PhysRevLett.54.2619} {\bibfield  {journal} {\bibinfo
  {journal} {Phys. Rev. Lett.}\ }\textbf {\bibinfo {volume} {54}},\ \bibinfo
  {pages} {2619} (\bibinfo {year} {1985})}\BibitemShut {NoStop}%
\bibitem [{\citenamefont {Brune}\ \emph {et~al.}(1994)\citenamefont {Brune},
  \citenamefont {R\"oder}, \citenamefont {Boragno},\ and\ \citenamefont
  {Kern}}]{Brune94}%
  \BibitemOpen
  \bibfield  {author} {\bibinfo {author} {\bibfnamefont {H.}~\bibnamefont
  {Brune}}, \bibinfo {author} {\bibfnamefont {H.}~\bibnamefont {R\"oder}},
  \bibinfo {author} {\bibfnamefont {C.}~\bibnamefont {Boragno}}, \ and\
  \bibinfo {author} {\bibfnamefont {K.}~\bibnamefont {Kern}},\ }\href {\doibase
  10.1103/PhysRevB.49.2997} {\bibfield  {journal} {\bibinfo  {journal} {Phys.
  Rev. B}\ }\textbf {\bibinfo {volume} {49}},\ \bibinfo {pages} {2997}
  (\bibinfo {year} {1994})}\BibitemShut {NoStop}%
\bibitem [{\citenamefont {Ling}\ \emph {et~al.}(2004)\citenamefont {Ling},
  \citenamefont {de~la Figuera}, \citenamefont {Bartelt}, \citenamefont
  {Hwang}, \citenamefont {Schmid}, \citenamefont {Thayer},\ and\ \citenamefont
  {Hamilton}}]{Lin04}%
  \BibitemOpen
  \bibfield  {author} {\bibinfo {author} {\bibfnamefont {W.~L.}\ \bibnamefont
  {Ling}}, \bibinfo {author} {\bibfnamefont {J.}~\bibnamefont {de~la Figuera}},
  \bibinfo {author} {\bibfnamefont {N.~C.}\ \bibnamefont {Bartelt}}, \bibinfo
  {author} {\bibfnamefont {R.~Q.}\ \bibnamefont {Hwang}}, \bibinfo {author}
  {\bibfnamefont {A.~K.}\ \bibnamefont {Schmid}}, \bibinfo {author}
  {\bibfnamefont {G.~E.}\ \bibnamefont {Thayer}}, \ and\ \bibinfo {author}
  {\bibfnamefont {J.~C.}\ \bibnamefont {Hamilton}},\ }\href {\doibase
  10.1103/PhysRevLett.92.116102} {\bibfield  {journal} {\bibinfo  {journal}
  {Phys. Rev. Lett.}\ }\textbf {\bibinfo {volume} {92}},\ \bibinfo {pages}
  {116102} (\bibinfo {year} {2004})}\BibitemShut {NoStop}%
\bibitem [{\citenamefont {Hwang}\ \emph {et~al.}(1995)\citenamefont {Hwang},
  \citenamefont {Hamilton}, \citenamefont {Stevens},\ and\ \citenamefont
  {Foiles}}]{Hwang95}%
  \BibitemOpen
  \bibfield  {author} {\bibinfo {author} {\bibfnamefont {R.~Q.}\ \bibnamefont
  {Hwang}}, \bibinfo {author} {\bibfnamefont {J.~C.}\ \bibnamefont {Hamilton}},
  \bibinfo {author} {\bibfnamefont {J.~L.}\ \bibnamefont {Stevens}}, \ and\
  \bibinfo {author} {\bibfnamefont {S.~M.}\ \bibnamefont {Foiles}},\ }\href
  {\doibase 10.1103/PhysRevLett.75.4242} {\bibfield  {journal} {\bibinfo
  {journal} {Phys. Rev. Lett.}\ }\textbf {\bibinfo {volume} {75}},\ \bibinfo
  {pages} {4242} (\bibinfo {year} {1995})}\BibitemShut {NoStop}%
\bibitem [{\citenamefont {G\"unther}\ \emph {et~al.}(1995)\citenamefont
  {G\"unther}, \citenamefont {Vrijmoeth}, \citenamefont {Hwang},\ and\
  \citenamefont {Behm}}]{Gunther95}%
  \BibitemOpen
  \bibfield  {author} {\bibinfo {author} {\bibfnamefont {C.}~\bibnamefont
  {G\"unther}}, \bibinfo {author} {\bibfnamefont {J.}~\bibnamefont
  {Vrijmoeth}}, \bibinfo {author} {\bibfnamefont {R.~Q.}\ \bibnamefont
  {Hwang}}, \ and\ \bibinfo {author} {\bibfnamefont {R.~J.}\ \bibnamefont
  {Behm}},\ }\href {\doibase 10.1103/PhysRevLett.74.754} {\bibfield  {journal}
  {\bibinfo  {journal} {Phys. Rev. Lett.}\ }\textbf {\bibinfo {volume} {74}},\
  \bibinfo {pages} {754} (\bibinfo {year} {1995})}\BibitemShut {NoStop}%
\bibitem [{\citenamefont {Butz}\ \emph {et~al.}(2014)\citenamefont {Butz},
  \citenamefont {Dolle}, \citenamefont {Niekiel}, \citenamefont {Weber},
  \citenamefont {Waldmann}, \citenamefont {Weber}, \citenamefont {Meyer},\ and\
  \citenamefont {Spiecker}}]{Butz14}%
  \BibitemOpen
  \bibfield  {author} {\bibinfo {author} {\bibfnamefont {B.}~\bibnamefont
  {Butz}}, \bibinfo {author} {\bibfnamefont {C.}~\bibnamefont {Dolle}},
  \bibinfo {author} {\bibfnamefont {F.}~\bibnamefont {Niekiel}}, \bibinfo
  {author} {\bibfnamefont {K.}~\bibnamefont {Weber}}, \bibinfo {author}
  {\bibfnamefont {D.}~\bibnamefont {Waldmann}}, \bibinfo {author}
  {\bibfnamefont {H.~B.}\ \bibnamefont {Weber}}, \bibinfo {author}
  {\bibfnamefont {B.}~\bibnamefont {Meyer}}, \ and\ \bibinfo {author}
  {\bibfnamefont {E.}~\bibnamefont {Spiecker}},\ }\href@noop {} {\bibfield
  {journal} {\bibinfo  {journal} {Nature}\ }\textbf {\bibinfo {volume} {505}},\
  \bibinfo {pages} {533} (\bibinfo {year} {2014})}\BibitemShut {NoStop}%
\bibitem [{\citenamefont {Schaff}\ \emph {et~al.}(2001)\citenamefont {Schaff},
  \citenamefont {Schmid}, \citenamefont {Bartelt}, \citenamefont {de~la
  Figuera},\ and\ \citenamefont {Hwang}}]{Schaff01}%
  \BibitemOpen
  \bibfield  {author} {\bibinfo {author} {\bibfnamefont {O.}~\bibnamefont
  {Schaff}}, \bibinfo {author} {\bibfnamefont {A.~K.}\ \bibnamefont {Schmid}},
  \bibinfo {author} {\bibfnamefont {N.~C.}\ \bibnamefont {Bartelt}}, \bibinfo
  {author} {\bibfnamefont {J.}~\bibnamefont {de~la Figuera}}, \ and\ \bibinfo
  {author} {\bibfnamefont {R.~Q.}\ \bibnamefont {Hwang}},\ }\href {\doibase
  https://doi.org/10.1016/S0921-5093(01)00977-7} {\bibfield  {journal}
  {\bibinfo  {journal} {Materials Science and Engineering: A}\ }\textbf
  {\bibinfo {volume} {319-321}},\ \bibinfo {pages} {914 } (\bibinfo {year}
  {2001})}\BibitemShut {NoStop}%
\bibitem [{\citenamefont {Wang}\ \emph {et~al.}(2010)\citenamefont {Wang},
  \citenamefont {Han}, \citenamefont {Liu}, \citenamefont {Yue}, \citenamefont
  {Zhang},\ and\ \citenamefont {Ma}}]{Wang10}%
  \BibitemOpen
  \bibfield  {author} {\bibinfo {author} {\bibfnamefont {L.}~\bibnamefont
  {Wang}}, \bibinfo {author} {\bibfnamefont {X.}~\bibnamefont {Han}}, \bibinfo
  {author} {\bibfnamefont {P.}~\bibnamefont {Liu}}, \bibinfo {author}
  {\bibfnamefont {Y.}~\bibnamefont {Yue}}, \bibinfo {author} {\bibfnamefont
  {Z.}~\bibnamefont {Zhang}}, \ and\ \bibinfo {author} {\bibfnamefont
  {E.}~\bibnamefont {Ma}},\ }\href {\doibase 10.1103/PhysRevLett.105.135501}
  {\bibfield  {journal} {\bibinfo  {journal} {Phys. Rev. Lett.}\ }\textbf
  {\bibinfo {volume} {105}},\ \bibinfo {pages} {135501} (\bibinfo {year}
  {2010})}\BibitemShut {NoStop}%
\bibitem [{\citenamefont {Chauraud}\ \emph {et~al.}(2019)\citenamefont
  {Chauraud}, \citenamefont {Durinck}, \citenamefont {Drouet}, \citenamefont
  {Vernisse}, \citenamefont {Bonneville},\ and\ \citenamefont
  {Coupeau}}]{Chauraud19}%
  \BibitemOpen
  \bibfield  {author} {\bibinfo {author} {\bibfnamefont {D.}~\bibnamefont
  {Chauraud}}, \bibinfo {author} {\bibfnamefont {J.}~\bibnamefont {Durinck}},
  \bibinfo {author} {\bibfnamefont {M.}~\bibnamefont {Drouet}}, \bibinfo
  {author} {\bibfnamefont {L.}~\bibnamefont {Vernisse}}, \bibinfo {author}
  {\bibfnamefont {J.}~\bibnamefont {Bonneville}}, \ and\ \bibinfo {author}
  {\bibfnamefont {C.}~\bibnamefont {Coupeau}},\ }\href {\doibase
  10.1103/PhysRevB.99.195404} {\bibfield  {journal} {\bibinfo  {journal} {Phys.
  Rev. B}\ }\textbf {\bibinfo {volume} {99}},\ \bibinfo {pages} {195404}
  (\bibinfo {year} {2019})}\BibitemShut {NoStop}%
\bibitem [{\citenamefont {Th\"urmer}\ \emph {et~al.}(2004)\citenamefont
  {Th\"urmer}, \citenamefont {Carter}, \citenamefont {Bartelt},\ and\
  \citenamefont {Hwang}}]{Thurmer04}%
  \BibitemOpen
  \bibfield  {author} {\bibinfo {author} {\bibfnamefont {K.}~\bibnamefont
  {Th\"urmer}}, \bibinfo {author} {\bibfnamefont {C.~B.}\ \bibnamefont
  {Carter}}, \bibinfo {author} {\bibfnamefont {N.~C.}\ \bibnamefont {Bartelt}},
  \ and\ \bibinfo {author} {\bibfnamefont {R.~Q.}\ \bibnamefont {Hwang}},\
  }\href {\doibase 10.1103/PhysRevLett.92.106101} {\bibfield  {journal}
  {\bibinfo  {journal} {Phys. Rev. Lett.}\ }\textbf {\bibinfo {volume} {92}},\
  \bibinfo {pages} {106101} (\bibinfo {year} {2004})}\BibitemShut {NoStop}%
\bibitem [{\citenamefont {de~la Figuera}\ \emph {et~al.}(2003)\citenamefont
  {de~la Figuera}, \citenamefont {Carter}, \citenamefont {Bartelt},\ and\
  \citenamefont {Hwang}}]{Figuera03}%
  \BibitemOpen
  \bibfield  {author} {\bibinfo {author} {\bibfnamefont {J.}~\bibnamefont
  {de~la Figuera}}, \bibinfo {author} {\bibfnamefont {C.}~\bibnamefont
  {Carter}}, \bibinfo {author} {\bibfnamefont {N.}~\bibnamefont {Bartelt}}, \
  and\ \bibinfo {author} {\bibfnamefont {R.}~\bibnamefont {Hwang}},\ }\href
  {\doibase https://doi.org/10.1016/S0039-6028(03)00401-1} {\bibfield
  {journal} {\bibinfo  {journal} {Surface Science}\ }\textbf {\bibinfo {volume}
  {531}},\ \bibinfo {pages} {29} (\bibinfo {year} {2003})}\BibitemShut
  {NoStop}%
\bibitem [{\citenamefont {Fleurence}\ \emph {et~al.}(2012)\citenamefont
  {Fleurence}, \citenamefont {Friedlein}, \citenamefont {Ozaki}, \citenamefont
  {Kawai}, \citenamefont {Wang},\ and\ \citenamefont
  {Yamada-Takamura}}]{Fleurence12}%
  \BibitemOpen
  \bibfield  {author} {\bibinfo {author} {\bibfnamefont {A.}~\bibnamefont
  {Fleurence}}, \bibinfo {author} {\bibfnamefont {R.}~\bibnamefont
  {Friedlein}}, \bibinfo {author} {\bibfnamefont {T.}~\bibnamefont {Ozaki}},
  \bibinfo {author} {\bibfnamefont {H.}~\bibnamefont {Kawai}}, \bibinfo
  {author} {\bibfnamefont {Y.}~\bibnamefont {Wang}}, \ and\ \bibinfo {author}
  {\bibfnamefont {Y.}~\bibnamefont {Yamada-Takamura}},\ }\href {\doibase
  10.1103/PhysRevLett.108.245501} {\bibfield  {journal} {\bibinfo  {journal}
  {Phys. Rev. Lett.}\ }\textbf {\bibinfo {volume} {108}},\ \bibinfo {pages}
  {245501} (\bibinfo {year} {2012})}\BibitemShut {NoStop}%
\bibitem [{\citenamefont {Vogt}\ \emph {et~al.}(2012)\citenamefont {Vogt},
  \citenamefont {De~Padova}, \citenamefont {Quaresima}, \citenamefont {Avila},
  \citenamefont {Frantzeskakis}, \citenamefont {Asensio}, \citenamefont
  {Resta}, \citenamefont {Ealet},\ and\ \citenamefont {Le~Lay}}]{Vogt12}%
  \BibitemOpen
  \bibfield  {author} {\bibinfo {author} {\bibfnamefont {P.}~\bibnamefont
  {Vogt}}, \bibinfo {author} {\bibfnamefont {P.}~\bibnamefont {De~Padova}},
  \bibinfo {author} {\bibfnamefont {C.}~\bibnamefont {Quaresima}}, \bibinfo
  {author} {\bibfnamefont {J.}~\bibnamefont {Avila}}, \bibinfo {author}
  {\bibfnamefont {E.}~\bibnamefont {Frantzeskakis}}, \bibinfo {author}
  {\bibfnamefont {M.~C.}\ \bibnamefont {Asensio}}, \bibinfo {author}
  {\bibfnamefont {A.}~\bibnamefont {Resta}}, \bibinfo {author} {\bibfnamefont
  {B.}~\bibnamefont {Ealet}}, \ and\ \bibinfo {author} {\bibfnamefont
  {G.}~\bibnamefont {Le~Lay}},\ }\href {\doibase
  10.1103/PhysRevLett.108.155501} {\bibfield  {journal} {\bibinfo  {journal}
  {Phys. Rev. Lett.}\ }\textbf {\bibinfo {volume} {108}},\ \bibinfo {pages}
  {155501} (\bibinfo {year} {2012})}\BibitemShut {NoStop}%
\bibitem [{\citenamefont {\relax{L. Chen, C.-C. Liu, B. Feng , X. He, P. Cheng,
  Z. Ding, S. Meng, Y. Yao, and K. Wu}}(2012)}]{Chen12}%
  \BibitemOpen
  \bibfield  {author} {\bibinfo {author} {\bibnamefont {\relax{L. Chen, C.-C.
  Liu, B. Feng , X. He, P. Cheng, Z. Ding, S. Meng, Y. Yao, and K. Wu}}},\
  }\href@noop {} {\bibfield  {journal} {\bibinfo  {journal} {Phys. Rev. Lett.}\
  }\textbf {\bibinfo {volume} {109}},\ \bibinfo {pages} {056804} (\bibinfo
  {year} {2012})}\BibitemShut {NoStop}%
\bibitem [{\citenamefont {Meng}\ \emph {et~al.}(2013)\citenamefont {Meng},
  \citenamefont {Wang}, \citenamefont {Zhang}, \citenamefont {Du},
  \citenamefont {Wu}, \citenamefont {Li}, \citenamefont {Zhang}, \citenamefont
  {Li}, \citenamefont {Zhou}, \citenamefont {Hofer},\ and\ \citenamefont
  {Gao}}]{Ming13}%
  \BibitemOpen
  \bibfield  {author} {\bibinfo {author} {\bibfnamefont {L.}~\bibnamefont
  {Meng}}, \bibinfo {author} {\bibfnamefont {Y.}~\bibnamefont {Wang}}, \bibinfo
  {author} {\bibfnamefont {L.}~\bibnamefont {Zhang}}, \bibinfo {author}
  {\bibfnamefont {S.}~\bibnamefont {Du}}, \bibinfo {author} {\bibfnamefont
  {R.}~\bibnamefont {Wu}}, \bibinfo {author} {\bibfnamefont {L.}~\bibnamefont
  {Li}}, \bibinfo {author} {\bibfnamefont {Y.}~\bibnamefont {Zhang}}, \bibinfo
  {author} {\bibfnamefont {G.}~\bibnamefont {Li}}, \bibinfo {author}
  {\bibfnamefont {H.}~\bibnamefont {Zhou}}, \bibinfo {author} {\bibfnamefont
  {W.~A.}\ \bibnamefont {Hofer}}, \ and\ \bibinfo {author} {\bibfnamefont
  {H.-J.}\ \bibnamefont {Gao}},\ }\href@noop {} {\bibfield  {journal} {\bibinfo
   {journal} {Nano Letters}\ }\textbf {\bibinfo {volume} {13}},\ \bibinfo
  {pages} {685} (\bibinfo {year} {2013})}\BibitemShut {NoStop}%
\bibitem [{\citenamefont {Aizawa}\ \emph {et~al.}(2014)\citenamefont {Aizawa},
  \citenamefont {Suehara},\ and\ \citenamefont {Otani}}]{Aizawa14}%
  \BibitemOpen
  \bibfield  {author} {\bibinfo {author} {\bibfnamefont {T.}~\bibnamefont
  {Aizawa}}, \bibinfo {author} {\bibfnamefont {S.}~\bibnamefont {Suehara}}, \
  and\ \bibinfo {author} {\bibfnamefont {S.}~\bibnamefont {Otani}},\
  }\href@noop {} {\bibfield  {journal} {\bibinfo  {journal} {The Journal of
  Physical Chemistry C}\ }\textbf {\bibinfo {volume} {118}},\ \bibinfo {pages}
  {23049} (\bibinfo {year} {2014})}\BibitemShut {NoStop}%
\bibitem [{\citenamefont {Cahangirov}\ \emph {et~al.}(2009)\citenamefont
  {Cahangirov}, \citenamefont {Topsakal}, \citenamefont {Akt\"urk},
  \citenamefont {\ifmmode~\mbox{\c{S}}\else \c{S}\fi{}ahin},\ and\
  \citenamefont {Ciraci}}]{Cahangirov09}%
  \BibitemOpen
  \bibfield  {author} {\bibinfo {author} {\bibfnamefont {S.}~\bibnamefont
  {Cahangirov}}, \bibinfo {author} {\bibfnamefont {M.}~\bibnamefont
  {Topsakal}}, \bibinfo {author} {\bibfnamefont {E.}~\bibnamefont {Akt\"urk}},
  \bibinfo {author} {\bibfnamefont {H.}~\bibnamefont
  {\ifmmode~\mbox{\c{S}}\else \c{S}\fi{}ahin}}, \ and\ \bibinfo {author}
  {\bibfnamefont {S.}~\bibnamefont {Ciraci}},\ }\href {\doibase
  10.1103/PhysRevLett.102.236804} {\bibfield  {journal} {\bibinfo  {journal}
  {Phys. Rev. Lett.}\ }\textbf {\bibinfo {volume} {102}},\ \bibinfo {pages}
  {236804} (\bibinfo {year} {2009})}\BibitemShut {NoStop}%
\bibitem [{\citenamefont {Fleurence}\ and\ \citenamefont
  {Yamada-Takamura}(2017)}]{Fleurence17}%
  \BibitemOpen
  \bibfield  {author} {\bibinfo {author} {\bibfnamefont {A.}~\bibnamefont
  {Fleurence}}\ and\ \bibinfo {author} {\bibfnamefont {Y.}~\bibnamefont
  {Yamada-Takamura}},\ }\href {\doibase 10.1063/1.4974467} {\bibfield
  {journal} {\bibinfo  {journal} {Applied Physics Letters}\ }\textbf {\bibinfo
  {volume} {110}},\ \bibinfo {pages} {041601} (\bibinfo {year}
  {2017})}\BibitemShut {NoStop}%
\bibitem [{\citenamefont {Fleurence}\ \emph {et~al.}(2016)\citenamefont
  {Fleurence}, \citenamefont {Gill}, \citenamefont {Friedlein}, \citenamefont
  {Sadowski}, \citenamefont {Aoyagi}, \citenamefont {Copel}, \citenamefont
  {Tromp}, \citenamefont {Hirjibehedin},\ and\ \citenamefont
  {Yamada-Takamura}}]{Fleurence16}%
  \BibitemOpen
  \bibfield  {author} {\bibinfo {author} {\bibfnamefont {A.}~\bibnamefont
  {Fleurence}}, \bibinfo {author} {\bibfnamefont {T.~G.}\ \bibnamefont {Gill}},
  \bibinfo {author} {\bibfnamefont {R.}~\bibnamefont {Friedlein}}, \bibinfo
  {author} {\bibfnamefont {J.~T.}\ \bibnamefont {Sadowski}}, \bibinfo {author}
  {\bibfnamefont {K.}~\bibnamefont {Aoyagi}}, \bibinfo {author} {\bibfnamefont
  {M.}~\bibnamefont {Copel}}, \bibinfo {author} {\bibfnamefont {R.~M.}\
  \bibnamefont {Tromp}}, \bibinfo {author} {\bibfnamefont {C.~F.}\ \bibnamefont
  {Hirjibehedin}}, \ and\ \bibinfo {author} {\bibfnamefont {Y.}~\bibnamefont
  {Yamada-Takamura}},\ }\href {\doibase 10.1063/1.4945370} {\bibfield
  {journal} {\bibinfo  {journal} {Applied Physics Letters}\ }\textbf {\bibinfo
  {volume} {108}},\ \bibinfo {pages} {151902} (\bibinfo {year}
  {2016})}\BibitemShut {NoStop}%
\bibitem [{\citenamefont {Yamada-Takamura}\ \emph {et~al.}(2010)\citenamefont
  {Yamada-Takamura}, \citenamefont {Bussolotti}, \citenamefont {Fleurence},
  \citenamefont {Bera},\ and\ \citenamefont {Friedlein}}]{YYT10}%
  \BibitemOpen
  \bibfield  {author} {\bibinfo {author} {\bibfnamefont {Y.}~\bibnamefont
  {Yamada-Takamura}}, \bibinfo {author} {\bibfnamefont {F.}~\bibnamefont
  {Bussolotti}}, \bibinfo {author} {\bibfnamefont {A.}~\bibnamefont
  {Fleurence}}, \bibinfo {author} {\bibfnamefont {S.}~\bibnamefont {Bera}}, \
  and\ \bibinfo {author} {\bibfnamefont {R.}~\bibnamefont {Friedlein}},\ }\href
  {\doibase 10.1063/1.3481414} {\bibfield  {journal} {\bibinfo  {journal}
  {Applied Physics Letters}\ }\textbf {\bibinfo {volume} {97}},\ \bibinfo
  {pages} {073109} (\bibinfo {year} {2010})}\BibitemShut {NoStop}%
\bibitem [{\citenamefont {Kohn}\ and\ \citenamefont {Sham}(1965)}]{Kohn65}%
  \BibitemOpen
  \bibfield  {author} {\bibinfo {author} {\bibfnamefont {W.}~\bibnamefont
  {Kohn}}\ and\ \bibinfo {author} {\bibfnamefont {L.~J.}\ \bibnamefont
  {Sham}},\ }\href {\doibase 10.1103/PhysRev.140.A1133} {\bibfield  {journal}
  {\bibinfo  {journal} {Phys. Rev.}\ }\textbf {\bibinfo {volume} {140}},\
  \bibinfo {pages} {A1133} (\bibinfo {year} {1965})}\BibitemShut {NoStop}%
\bibitem [{\citenamefont {Perdew}\ \emph {et~al.}(1996)\citenamefont {Perdew},
  \citenamefont {Burke},\ and\ \citenamefont {Ernzerhof}}]{Perdew96}%
  \BibitemOpen
  \bibfield  {author} {\bibinfo {author} {\bibfnamefont {J.~P.}\ \bibnamefont
  {Perdew}}, \bibinfo {author} {\bibfnamefont {K.}~\bibnamefont {Burke}}, \
  and\ \bibinfo {author} {\bibfnamefont {M.}~\bibnamefont {Ernzerhof}},\ }\href
  {\doibase 10.1103/PhysRevLett.77.3865} {\bibfield  {journal} {\bibinfo
  {journal} {Phys. Rev. Lett.}\ }\textbf {\bibinfo {volume} {77}},\ \bibinfo
  {pages} {3865} (\bibinfo {year} {1996})}\BibitemShut {NoStop}%
\bibitem [{\citenamefont {Ozaki}(2003)}]{Ozaki03}%
  \BibitemOpen
  \bibfield  {author} {\bibinfo {author} {\bibfnamefont {T.}~\bibnamefont
  {Ozaki}},\ }\href {\doibase 10.1103/PhysRevB.67.155108} {\bibfield  {journal}
  {\bibinfo  {journal} {Phys. Rev. B}\ }\textbf {\bibinfo {volume} {67}},\
  \bibinfo {pages} {155108} (\bibinfo {year} {2003})}\BibitemShut {NoStop}%
\bibitem [{Oza()}]{Ozakiwebsite}%
  \BibitemOpen
  \href@noop {} {}\bibinfo {note} {Http://www.openmx-square.org/}\BibitemShut
  {NoStop}%
\bibitem [{SM()}]{SM}%
  \BibitemOpen
  \href@noop {} {}\bibinfo {note} {Supplementary materials}\BibitemShut
  {NoStop}%
\bibitem [{\citenamefont {Lee}\ \emph {et~al.}(2013)\citenamefont {Lee},
  \citenamefont {Fleurence}, \citenamefont {Friedlein}, \citenamefont
  {Yamada-Takamura},\ and\ \citenamefont {Ozaki}}]{Lee13}%
  \BibitemOpen
  \bibfield  {author} {\bibinfo {author} {\bibfnamefont {C.-C.}\ \bibnamefont
  {Lee}}, \bibinfo {author} {\bibfnamefont {A.}~\bibnamefont {Fleurence}},
  \bibinfo {author} {\bibfnamefont {R.}~\bibnamefont {Friedlein}}, \bibinfo
  {author} {\bibfnamefont {Y.}~\bibnamefont {Yamada-Takamura}}, \ and\ \bibinfo
  {author} {\bibfnamefont {T.}~\bibnamefont {Ozaki}},\ }\href {\doibase
  10.1103/PhysRevB.88.165404} {\bibfield  {journal} {\bibinfo  {journal} {Phys.
  Rev. B}\ }\textbf {\bibinfo {volume} {88}},\ \bibinfo {pages} {165404}
  (\bibinfo {year} {2013})}\BibitemShut {NoStop}%
\bibitem [{\citenamefont {Lee}\ \emph {et~al.}(2014)\citenamefont {Lee},
  \citenamefont {Fleurence}, \citenamefont {Yamada-Takamura}, \citenamefont
  {Ozaki},\ and\ \citenamefont {Friedlein}}]{Lee14ARPES}%
  \BibitemOpen
  \bibfield  {author} {\bibinfo {author} {\bibfnamefont {C.-C.}\ \bibnamefont
  {Lee}}, \bibinfo {author} {\bibfnamefont {A.}~\bibnamefont {Fleurence}},
  \bibinfo {author} {\bibfnamefont {Y.}~\bibnamefont {Yamada-Takamura}},
  \bibinfo {author} {\bibfnamefont {T.}~\bibnamefont {Ozaki}}, \ and\ \bibinfo
  {author} {\bibfnamefont {R.}~\bibnamefont {Friedlein}},\ }\href {\doibase
  10.1103/PhysRevB.90.075422} {\bibfield  {journal} {\bibinfo  {journal} {Phys.
  Rev. B}\ }\textbf {\bibinfo {volume} {90}},\ \bibinfo {pages} {075422}
  (\bibinfo {year} {2014})}\BibitemShut {NoStop}%
\bibitem [{\citenamefont {Warner}\ \emph {et~al.}(2017)\citenamefont {Warner},
  \citenamefont {Gill}, \citenamefont {Caciuc}, \citenamefont {Atodiresei},
  \citenamefont {Fleurence}, \citenamefont {Yoshida}, \citenamefont {Hasegawa},
  \citenamefont {Bl\"ugel}, \citenamefont {Yamada-Takamura},\ and\
  \citenamefont {Hirjibehedin}}]{Warner17}%
  \BibitemOpen
  \bibfield  {author} {\bibinfo {author} {\bibfnamefont {B.}~\bibnamefont
  {Warner}}, \bibinfo {author} {\bibfnamefont {T.~G.}\ \bibnamefont {Gill}},
  \bibinfo {author} {\bibfnamefont {V.}~\bibnamefont {Caciuc}}, \bibinfo
  {author} {\bibfnamefont {N.}~\bibnamefont {Atodiresei}}, \bibinfo {author}
  {\bibfnamefont {A.}~\bibnamefont {Fleurence}}, \bibinfo {author}
  {\bibfnamefont {Y.}~\bibnamefont {Yoshida}}, \bibinfo {author} {\bibfnamefont
  {Y.}~\bibnamefont {Hasegawa}}, \bibinfo {author} {\bibfnamefont
  {S.}~\bibnamefont {Bl\"ugel}}, \bibinfo {author} {\bibfnamefont
  {Y.}~\bibnamefont {Yamada-Takamura}}, \ and\ \bibinfo {author} {\bibfnamefont
  {C.~F.}\ \bibnamefont {Hirjibehedin}},\ }\href {\doibase
  10.1002/adma.201703929} {\bibfield  {journal} {\bibinfo  {journal} {Advanced
  Materials}\ }\textbf {\bibinfo {volume} {29}},\ \bibinfo {pages} {1703929}
  (\bibinfo {year} {2017})}\BibitemShut {NoStop}%
\bibitem [{\citenamefont {Nogami}\ \emph {et~al.}(2019)\citenamefont {Nogami},
  \citenamefont {Fleurence}, \citenamefont {Yamada-Takamura},\ and\
  \citenamefont {Tomitori}}]{Nogami19}%
  \BibitemOpen
  \bibfield  {author} {\bibinfo {author} {\bibfnamefont {M.}~\bibnamefont
  {Nogami}}, \bibinfo {author} {\bibfnamefont {A.}~\bibnamefont {Fleurence}},
  \bibinfo {author} {\bibfnamefont {Y.}~\bibnamefont {Yamada-Takamura}}, \ and\
  \bibinfo {author} {\bibfnamefont {M.}~\bibnamefont {Tomitori}},\ }\href@noop
  {} {\bibfield  {journal} {\bibinfo  {journal} {Advanced Materials
  Interfaces}\ }\textbf {\bibinfo {volume} {6}},\ \bibinfo {pages} {1801278}
  (\bibinfo {year} {2019})}\BibitemShut {NoStop}%
\bibitem [{\citenamefont {Gong}\ \emph {et~al.}(2015)\citenamefont {Gong},
  \citenamefont {Robertson}, \citenamefont {He}, \citenamefont {Lee},
  \citenamefont {Yoon}, \citenamefont {Allen}, \citenamefont {Kirkland},\ and\
  \citenamefont {Warner}}]{Gong15}%
  \BibitemOpen
  \bibfield  {author} {\bibinfo {author} {\bibfnamefont {C.}~\bibnamefont
  {Gong}}, \bibinfo {author} {\bibfnamefont {A.~W.}\ \bibnamefont {Robertson}},
  \bibinfo {author} {\bibfnamefont {K.}~\bibnamefont {He}}, \bibinfo {author}
  {\bibfnamefont {G.-D.}\ \bibnamefont {Lee}}, \bibinfo {author} {\bibfnamefont
  {E.}~\bibnamefont {Yoon}}, \bibinfo {author} {\bibfnamefont {C.~S.}\
  \bibnamefont {Allen}}, \bibinfo {author} {\bibfnamefont {A.~I.}\ \bibnamefont
  {Kirkland}}, \ and\ \bibinfo {author} {\bibfnamefont {J.~H.}\ \bibnamefont
  {Warner}},\ }\href {\doibase 10.1021/acsnano.5b05355} {\bibfield  {journal}
  {\bibinfo  {journal} {ACS Nano}\ }\textbf {\bibinfo {volume} {9}},\ \bibinfo
  {pages} {10066} (\bibinfo {year} {2015})}\BibitemShut {NoStop}%
\bibitem [{\citenamefont {Warner}\ \emph {et~al.}(2012)\citenamefont {Warner},
  \citenamefont {Margine}, \citenamefont {Mukai}, \citenamefont {Robertson},
  \citenamefont {Giustino},\ and\ \citenamefont {Kirkland}}]{Warner12}%
  \BibitemOpen
  \bibfield  {author} {\bibinfo {author} {\bibfnamefont {J.~H.}\ \bibnamefont
  {Warner}}, \bibinfo {author} {\bibfnamefont {E.~R.}\ \bibnamefont {Margine}},
  \bibinfo {author} {\bibfnamefont {M.}~\bibnamefont {Mukai}}, \bibinfo
  {author} {\bibfnamefont {A.~W.}\ \bibnamefont {Robertson}}, \bibinfo {author}
  {\bibfnamefont {F.}~\bibnamefont {Giustino}}, \ and\ \bibinfo {author}
  {\bibfnamefont {A.~I.}\ \bibnamefont {Kirkland}},\ }\href {\doibase
  10.1126/science.1217529} {\bibfield  {journal} {\bibinfo  {journal}
  {Science}\ }\textbf {\bibinfo {volume} {337}},\ \bibinfo {pages} {209}
  (\bibinfo {year} {2012})}\BibitemShut {NoStop}%
\bibitem [{\citenamefont {Lehtinen}\ \emph {et~al.}(2013)\citenamefont
  {Lehtinen}, \citenamefont {Kurasch}, \citenamefont {Krasheninnikov},\ and\
  \citenamefont {Kaiser}}]{Lehtinen13}%
  \BibitemOpen
  \bibfield  {author} {\bibinfo {author} {\bibfnamefont {O.}~\bibnamefont
  {Lehtinen}}, \bibinfo {author} {\bibfnamefont {S.}~\bibnamefont {Kurasch}},
  \bibinfo {author} {\bibfnamefont {A.~V.}\ \bibnamefont {Krasheninnikov}}, \
  and\ \bibinfo {author} {\bibfnamefont {U.}~\bibnamefont {Kaiser}},\
  }\href@noop {} {\bibfield  {journal} {\bibinfo  {journal} {Nature
  Communications}\ }\textbf {\bibinfo {volume} {4}},\ \bibinfo {pages} {2098}
  (\bibinfo {year} {2013})}\BibitemShut {NoStop}%
\end{thebibliography}



%

\newpage

\newpage
\newpage

\newpage

\newpage
 
 \begin{table}[tbh]
\begin{tabular}{||c|c|c|c||}
\hline
\hline
 Domains sequence & Burgers vectors & $N_{Si}$   & E per Si atom (meV)  \\
\hline
\hline
$abcda$  & $\vec{b}_+ \times 2$, $\vec{b}_- \times 2 $ &104 &  0   \\
$abcda$, Adatoms at the boundaries  & $\vec{b}_+ \times 2$, $\vec{b}_- \times 2 $ & 108 &  66.9   \\
$abcda$, Adatoms at the domains & $\vec{b}_+ \times 2$, $\vec{b}_- \times 2 $ & 108 &  65.0   \\
$a$  & none &  108    & 30.5   \\
$aa$  &   $2(\vec{b}_{+}+\vec{b}_{-})=a_{Si}\vec{u}_x $ &104 &  15.8   \\
$acda$  & $\vec{b}_{+} +\vec{b}_-$, $\vec{b}_+$, $\vec{b}_-$  & 104 &  3.8   \\
$ada$  & $2\vec{b}_{+} +\vec{b}_-$,  $\vec{b}_-$ & 104 &  7.9   \\
$ada$  &   $-\vec{b}_-$, $\vec{b}_-$ & 108 &  39.7  \\

\hline
\hline

\end{tabular}
\caption{\textbf{Energy per Si atom in the 4 domains supercell for different domain sequences}. E is calculated by $E=\frac{E(ZrB_2/silicene)-E(ZrB_2)}{N_{Si}}$, where $E(ZrB_2/silicene)$ and $E(ZrB_2)$ are respectively  the energies of the slabs with and without silicene and $N _{Si}$ is the number of Si atoms. The  number and Burgers vectors of the dislocations in the supercell are indicated. The corresponding structures can be seen in Fig. S1 of supplementary materials \cite{SM}.}
\label{Table}
\end{table}

 \end{document}